\documentclass{article}

\usepackage{arxiv}

\usepackage[utf8]{inputenc} 
\usepackage[T1]{fontenc}    
\usepackage{hyperref}       
\usepackage{url}            
\usepackage{booktabs}       
\usepackage{amsfonts}       
\usepackage{nicefrac}       
\usepackage{microtype}      
\usepackage{lipsum}
\usepackage{graphicx}
\graphicspath{ {./images/} }
\usepackage{dblfloatfix}
\usepackage{amsmath}
\usepackage{xcolor}

\title{3DGS-to-PC: Convert a 3D Gaussian Splatting Scene into a Dense Point Cloud or Mesh}

\author{
 Lewis A G Stuart \\
  School of Computer Science\\
  University of Nottingham\\
  Jubilee Campus, Nottingham, NG8 1BB \\
  \texttt{lewis.stuart@nottingham.ac.uk} \\
  \And
 Michael P Pound\\
  School of Computer Science\\
  University of Nottingham\\
  Jubilee Campus, Nottingham, NG8 1BB \\
  \texttt{michael.pound@nottingham.ac.uk}  \\
}

\begin{document}
\maketitle
\begin{abstract}

3D Gaussian Splatting (3DGS) excels at producing highly detailed 3D reconstructions, but these scenes often require specialised renderers for effective visualisation. In contrast, point clouds are a widely used 3D representation and are compatible with most popular 3D processing software, yet converting 3DGS scenes into point clouds is a complex challenge.  In this work we introduce 3DGS-to-PC, a flexible and highly customisable framework that is capable of transforming 3DGS scenes into dense, high-accuracy point clouds. We sample points probabilistically from each Gaussian as a 3D density function. We additionally threshold new points using the Mahalanobis distance to the Gaussian centre, preventing extreme outliers. The result is a point cloud that closely represents the shape encoded into the 3D Gaussian scene. Individual Gaussians use spherical harmonics to adapt colours depending on view, and each point may contribute only subtle colour hints to the resulting rendered scene. To avoid spurious or incorrect colours that do not fit with the final point cloud, we recalculate Gaussian colours via a customised image rendering approach, assigning each Gaussian the colour of the pixel to which it contributes most across all views. 3DGS-to-PC also supports mesh generation through Poisson Surface Reconstruction, applied to points sampled from predicted surface Gaussians. This allows coloured meshes to be generated from 3DGS scenes without the need for re-training. This package is highly customisable and capability of simple integration into existing 3DGS pipelines. 3DGS-to-PC provides a powerful tool for converting 3DGS data into point cloud and surface-based formats. The project repository can be found at \textcolor{blue}{\href{https://github.com/Lewis-Stuart-11/3DGS-to-PC}{https://github.com/Lewis-Stuart-11/3DGS-to-PC}}.

\end{abstract}


\begin{figure*}[bp]
\centering
    \includegraphics[width=0.87\textwidth]{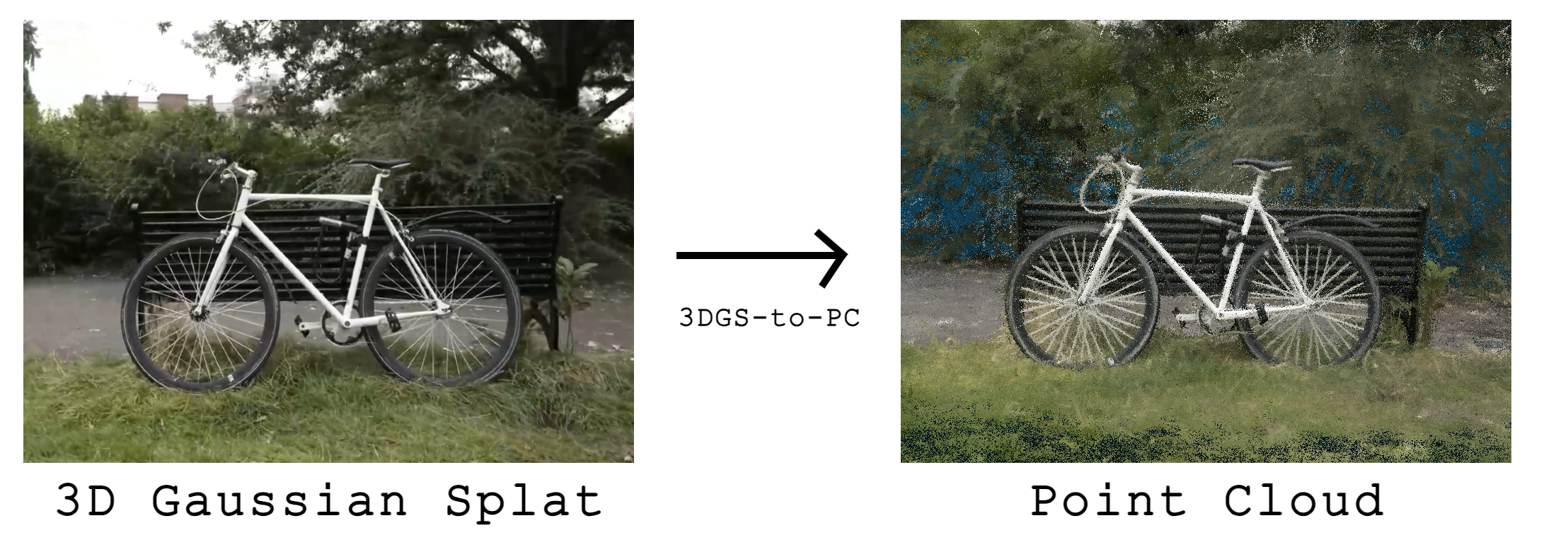}
     \caption{Showcase of how 3DGS-to-PC can convert the bike scene from the Mip-NeRF 360 \cite{abs-2111-12077} dataset from a 3D Gaussian splat into a dense point cloud.}
    \label{fig:resultsShowcase}
\end{figure*}

\section{Introduction}

3D Gaussian Splatting (3DGS) \cite{kerbl20233d} is a recent and popular approach to view synthesis. This method produces high-quality 3D representations from 2D images and their associated camera poses. However, visualising these representations requires specialised Gaussian renderers in order to display the 3D scene. In contrast, point clouds are a well-established 3D representation, wherein the objects in a scene are encoded as a series of discrete points, often with an associated colour. Point clouds have become a common part of many 3D pipelines, for example, as a step towards a full mesh representation \cite{huang2024surface}, as an input into deep learning frameworks \cite{PointNet, pointcloudSurvey}, or for scientific measurement \cite{lin2017measurement, agronomy9100596}. Furthermore, point clouds are compatible with numerous established 3D processing software. Nevertheless, 3DGS offer several compelling advantages over traditional point clouds. This includes a more accurate representation of 3D shape, with Gaussians containing 3D size and rotation information, and the ability to render scenes with greater detail. Although 3D Gaussian scenes can be loaded into point cloud software, such as CloudCompare \cite{girardeau2016cloudcompare}, only the centre of each Gaussians is utilised, neglecting geometric properties such as rotation and scale, which are critical for effectively representing areas of the scene. The result is that exported point clouds are more sparse than desired, and represent the shape of the underlying scene objects less accurately. \newline
Packages exist that can convert a 3DGS scene into a mesh, which can then be converted into a point cloud. However, these methods typically focus on only generating surface meshes during the training process, rather than providing a mechanism for converting the entire scene post-training. To address this gap, we introduce 3DGS-to-PC, a framework designed to convert a 3D Gaussian Scene into a dense point cloud. The framework accepts 3DGS data in .ply or .splat file formats, as well as camera poses in .json or COLMAP \cite{schoenberger2016sfm, schoenberger2016mvs} project formats. Users can specify the number of points that are generated, with a variety of additional customisation options to control point sampling. The process samples points probabilistically and evenly across the scene, according to the size of each Gaussian. Rendering of the scene is simultaneously used to colour each output point based on its contribution to the synthesised images, rather than its raw colour of the Gaussian it was sampled from. The resulting output consists of a coloured point cloud, stored as a .ply file, that accurately reflects the original Gaussian scene and is fully compatible with established point cloud processing tools. In this technical report, we provide details of the methodologies employed in 3DGS-to-PC, specifically: 

\begin{enumerate}
    \item The process of rendering new point colours to correct the issue of erroneous colours from sampling directly from each Gaussian colours.
    \item The optimised sampling of points across the 3D Gaussian scene.
    \item Our approach for converting Gaussian surfaces into a basic mesh representation.
\end{enumerate}

The project Github repository can be found at \textcolor{blue}{\href{https://github.com/Lewis-Stuart-11/3DGS-to-PC}{https://github.com/Lewis-Stuart-11/3DGS-to-PC.}}

\section{3DGS-to-PC}

\subsection{Initialising Gaussians}
Gaussians are loaded into the framework from either .splat or .ply files. To ensure the covariances of each Gaussian are positive-definite, we use regularisation techniques to ensure that points can be correctly sampled from each distribution. Each Gaussian is defined by a position, scale, rotation, opacity and spherical harmonics, which accurately determine the locations and rendering properties of each Gaussian in the scene. The covariance matrix representing each Gaussian is computed based on the rotation and scale parameters. Only the 0th degree of the spherical harmonics is used to determine the Gaussian colours, since rendering new Gaussians colours per camera from the spherical harmonics had a significant impact on rendering time. In large-scale scenes, the desired point cloud may represent only a subset of the full set of loaded Gaussians. Hence, we offer several additional filtering options to reduce the number of Gaussians in the scene prior to point generation. These include a bounding box removal, filtering of large Gaussians, and removing Gaussians with low opacity. 3DGS is optimised for rendering quality rather than object shape, and these filters can help reduce noise and improve the final point cloud quality.\newline

\subsection{Rendering Colours}
\label{sec:rendering_colours}

\begin{figure*}[h]
\centering
    \includegraphics[width=0.85\textwidth]{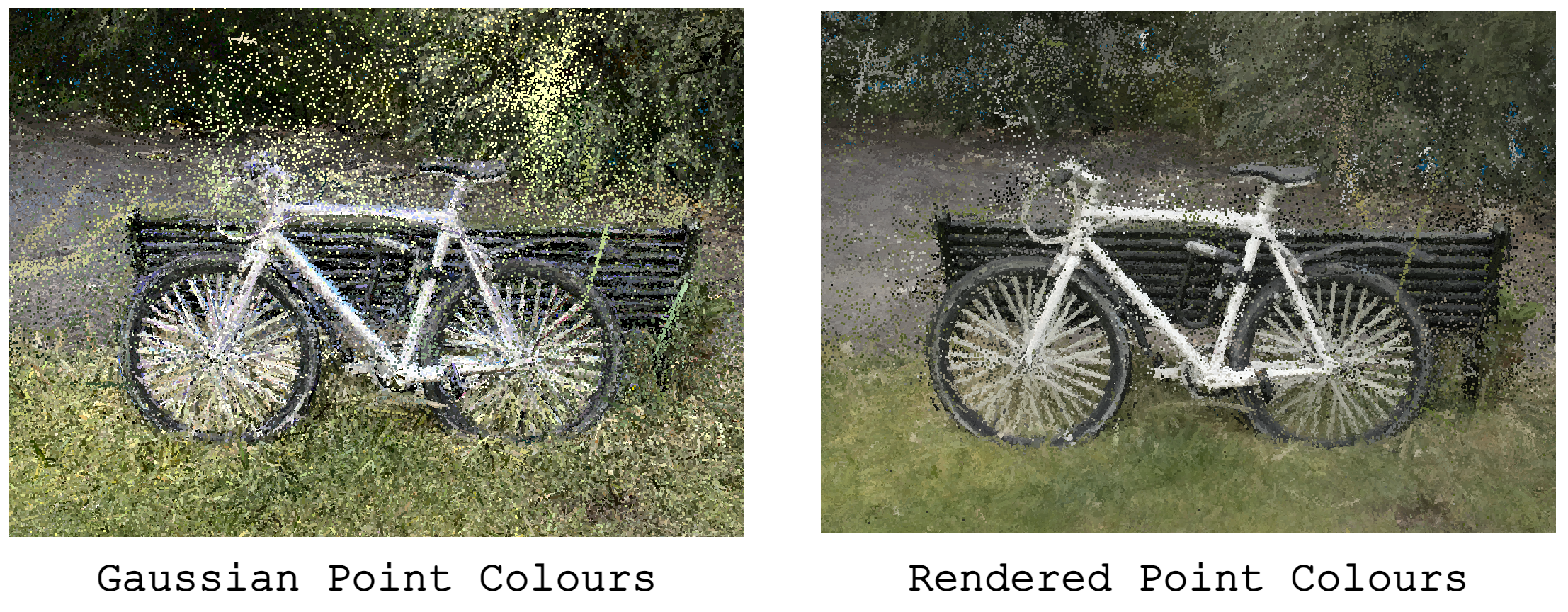}
     \caption{Comparison between the different techniques for generating point clouds of the Mip-NeRF 360 bike scene. The left shows a point cloud where the colours of each point are the colours of the Gaussians that they have been generated from. The right shows a point cloud where the colours have been generated via our colour rendering process. These point cloud are shown in the CloudCompare viewer, and all points were set to a size of 3 to show the effect of the noise produced from using colours from the original Gaussians.}
    \label{fig:colourComparison}
\end{figure*}
Sampling points from Gaussians in 3D space is a straightforward process, since the Gaussians represent 3D probability distribution functions, and new point positions can be produced at random according to this distribution. However, the challenge lies in accurately determining the colour of each of the sampled points. Directly assigning the Gaussian colours to each sampled point produces colours that fail to accurately reflect the scene, as illustrated in Figure \ref{fig:colourComparison}. This inaccuracy arises since, during rendering, multiple Gaussians in the scene are blended together to generate final pixel values. The contribution of a single Gaussian will be determined based on its position, the view direction, their opacity, and other Gaussians in that location. Each Gaussian may hold different and distinct colours that produce accurate pixel colours when rendered, but individually do not fit the colours of the surrounding environment. \newline
We address this issue by assigning colours to Gaussians based on how they contribute to the pixel colours in the rendered scene images. This process requires the original camera poses used during training or evaluation of the scene; to accurately render images of the 3DGS scene, a list of valid camera poses are required. These camera transforms are loaded as either .bin/.txt files, which are generated from COLMAP, or .json files structured according to the original NeRF \cite{mildenhall2020nerf} format. \newline 
For rendering the image of the scene, a custom Pytorch~\cite{paszke2019pytorchimperativestylehighperformance} renderer was developed, based on the Torch-Splatting \cite{torchsplatting} library, which is a pure Python implementation of the original 3DGS framework. The renderer was modified to dynamically adjust the tile size for each subsection of the image during rendering, in order to optimise the number of Gaussians per tile as a balance between efficiency and speed. If the number of Gaussians within a tile will exceed the GPU's memory capacity, the tile is subdivided into four smaller tiles. This subdivision process continues until all tiles can be rendered without exceeding memory limits. This approach ensures optimal parallelism and rendering speed while avoiding memory overflows. \newline
For each rendered pixel, we calculate the contribution that each Gaussian in the tile had to calculating the pixel colour. The pixel to which a Gaussian contributes the most to is then assigned as the new colour for that Gaussian. This contribution is determined based on the Gaussian’s position along the Z-axis in the image space and its opacity during rendering, as defined in Equation \ref{eq:contribution}. \newline
\begin{equation}\label{eq:contribution}
\Large  C_i = a_it_i,
\vspace{0.75em}
\end{equation}
where $i$ represents the Gaussians currently in the tile, $a$ is the alpha value of each Gaussian for this image, and $t$ is the current transmittance of Gaussian in the image space. The transmittance is calculated by taking the cumulative product of the current Gaussians in the tile, which are ordered based on each position along the Z-axis. Gaussians occluded by opaque Gaussians will have a much lower contribution than Gaussians closer to the current camera. \newline 
\begin{figure*}[t]
\centering
    \includegraphics[width=0.9\textwidth]{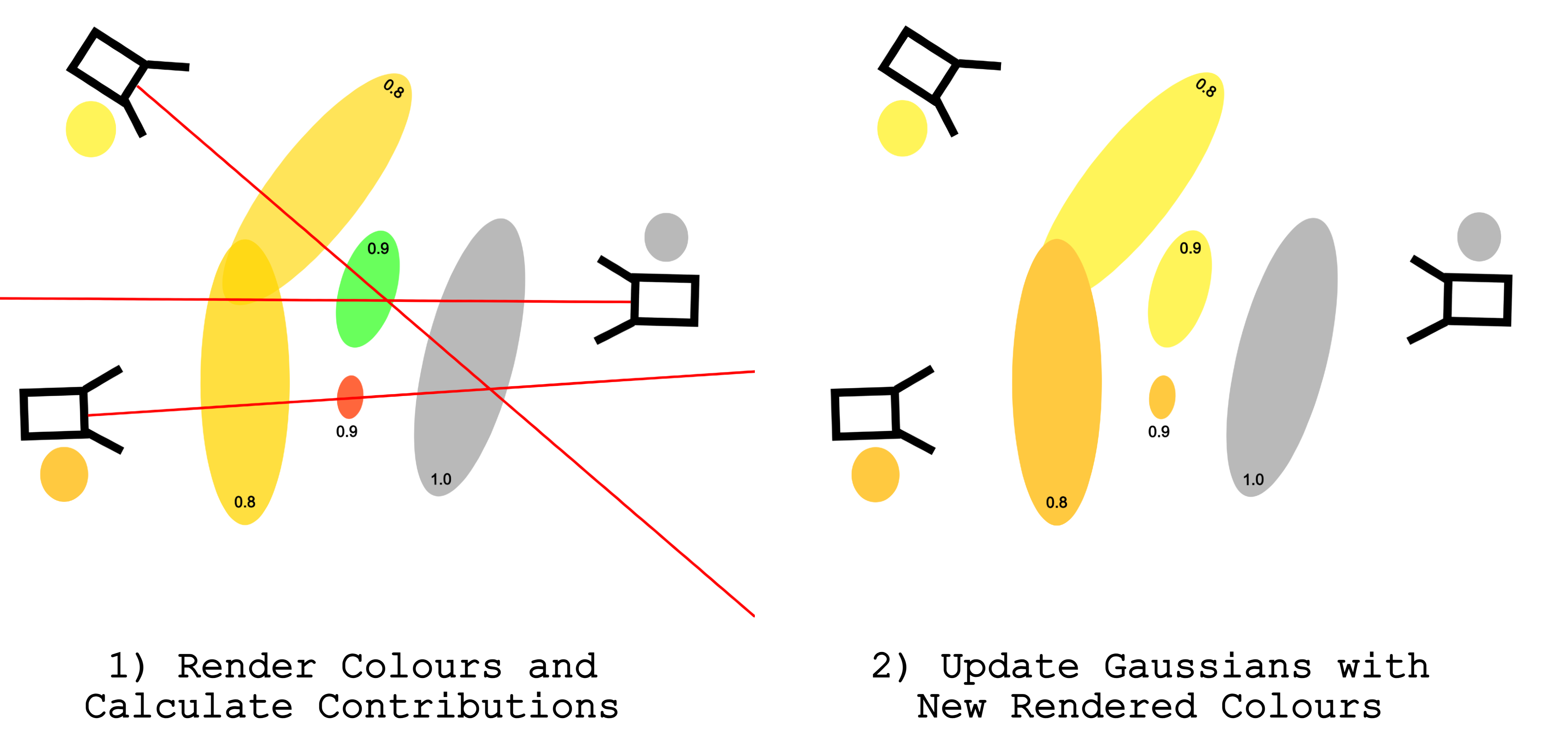}
     \caption{Demonstration of how the Gaussian colours are determined based on rendered colours in a scene. The scene consists of a set of Gaussians that make up the surface of a yellow object, with a seperate gray background surface. The transparency (alpha) values of each Gaussian are shown. The left side of the diagram demonstrates pixel colour rendering, where occluded Gaussians (e.g. green and red) contribute collectively to the final pixel colour, even though individually these do not match the environment. The right side illustrates the update process, where each Gaussian's colour is adjusted to match the pixel it contributed to the most. For instance, the green Gaussian is updated to reflect the yellow pixel rather than the gray one, since the gray Gaussian is opaque. In practice, this process involves more cameras and pixels than depicted in the simplified diagram.}
    \label{fig:renderingColoursShowcase}
\end{figure*}
An illustration of how this process produces accurate Gaussian colours is provided in Figure \ref{fig:renderingColoursShowcase}. The process begins by initialising the largest contribution of each Gaussian to zero and setting the Gaussian’s colour to the default background. For each rendered pixel in each image, the percentage contribution of each Gaussian to the final pixel colour is calculated, using Equation  \ref{eq:contribution}. If a Gaussian’s contribution to a pixel exceeds its current recorded largest contribution, the Gaussian’s colour is updated to match the pixel colour, and its largest contribution is updated accordingly. This process is repeated for every rendered image in the provided camera poses. \newline 
This approach assigns more accurate colours that reflect their appearance in the rendered scene, while occluded or less visible Gaussians adopt colours consistent with the visible surface Gaussians at that location. Gaussians that have a contribution value of zero, meaning that they have not been rendered in any of the provided images, are removed prior to point sampling. This step optimises the resulting point cloud by excluding redundant Gaussians. 

\subsection{Sampling Points}
\label{sec:sampling_points}

To convert each Gaussian into a series of points, its volume must first be determined, to ensure that the correct number of points can be assigned based on each Gaussian's size. The volume is calculated directly from the Gaussian’s scale using Equation \ref{eq:gaussian_volume}. 
\vspace{2em}
\begin{equation}\label{eq:gaussian_volume}
\large V =\sqrt{\sum_{i} \left( e^\text{s}_i \right)^2},
\vspace{1em}
\end{equation}
where $s$ is the scale values a Gaussian and $i$ represent each of the dimensions. The volume is calculated from the natural exponential of each dimension, such that that elongated Gaussians, which typically resemble large surfaces, have a larger volume.  Other methods exist for calculating the volume of a Gaussian, however we found empirically that this method produces the best distribution of points between Gaussians of different sizes. With the volumes established, new points are distributed to each Gaussian based on its relative volume, such that the total number of points generated across all Gaussians matches the requested point cloud size. This allows the user to generate very dense or more sparse point clouds as preferred. This approach ensures that larger Gaussians, which typically represent significant surfaces in the scene, are allocated more points. \newline
\begin{figure*}[t]
\centering
    \includegraphics[width=0.9\textwidth]{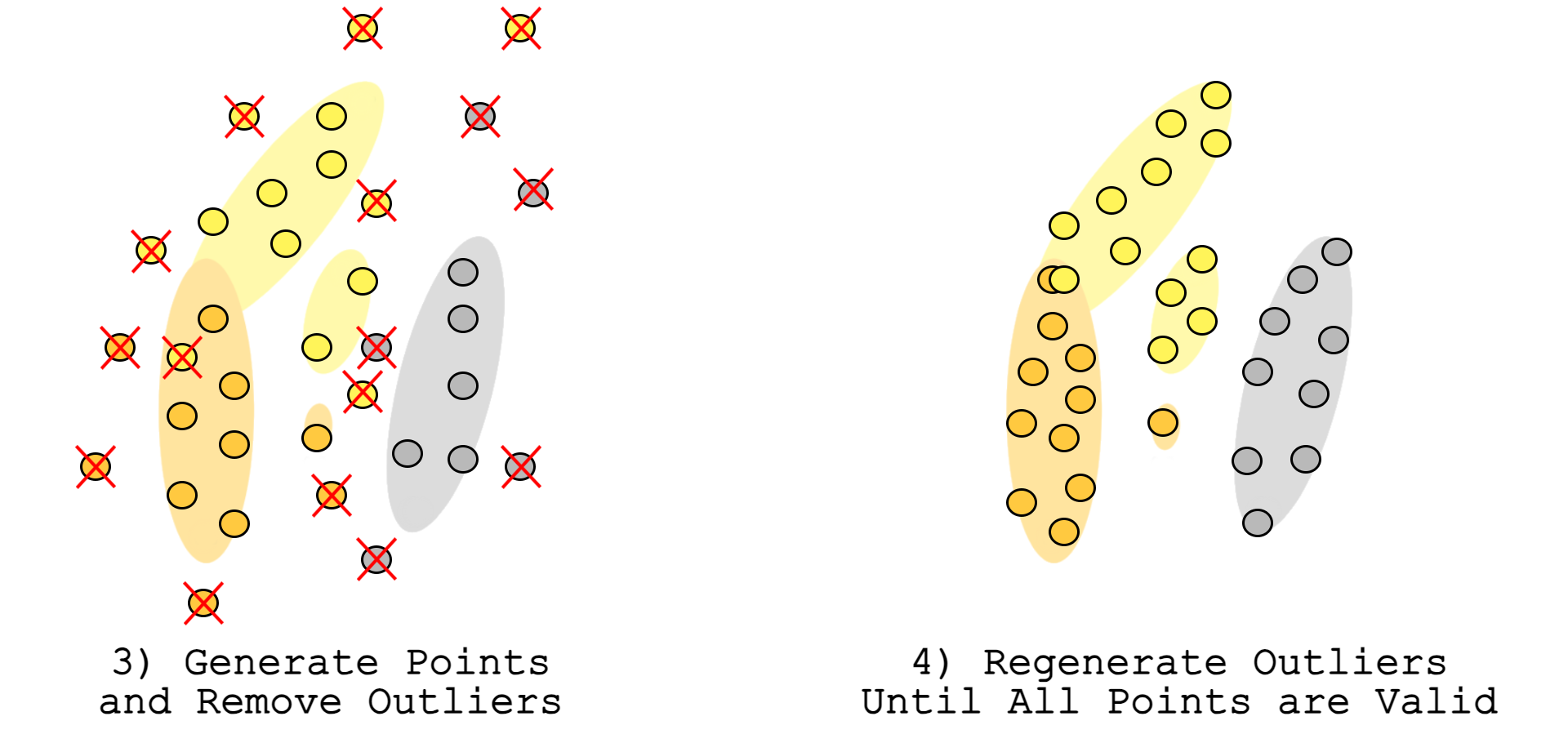}
     \caption{Demonstration of point generation from Gaussians with rendered colours. The number of points generated for each Gaussian is determined by its volume, calculated using Equation \ref{eq:gaussian_volume}, which is shown on the left of the diagram. Distances between the Gaussian centre and generated points are evaluated using Equation \ref{eq:mahalanobis}. Points exceeding a distance of 2 standard deviation are removed and regenerated, as shown on the right of the diagram. This iterative process ensures that the final point cloud accurately represents the scene, with the number of points per Gaussian being proportional to its volume. }
    \label{fig:samplingPointsShowcase}
\end{figure*}
The correct number of individual points are sampled from the Gaussian's multivariate normal distribution. However, this sampling process can only be performed in parallel with Gaussians that have the same number of points to sample. When the number of points in the point cloud is set to a large value, the range of the number of points to generate for each Gaussian will be extensive, making this sampling process slow. To address this, Gaussians are grouped together into bins based on the number of points to sample, with Gaussians requiring more than 50 points grouped into bins of 5-point intervals. This approach significantly increased the speed of the point generation process. If a precise number of points is required, this binning process can be bypassed, allowing for exact point sampling. However, in most cases, the binning strategy provides a practical balance between accuracy and computational efficiency. \newline
Because Gaussians represent probability distributions, there exists a chance for any individual point that it may appear far away from a Gaussian centre. To address this, the Mahalanobis distance between each generated point and its associated Gaussian is calculated, as defined in Equation \ref{eq:mahalanobis}.
\vspace{2em}
\begin{equation}\label{eq:mahalanobis}
\large D_M(\mathbf{x}) = \sqrt{(\mathbf{x} - \boldsymbol{\mu})^\top \mathbf{\Sigma}^{-1} (\mathbf{x} - \boldsymbol{\mu})},
\vspace{1em}
\end{equation}
where $x$ is the current point, $\boldsymbol{\mu}$ is the mean and $\boldsymbol{\Sigma}$ is the positive semi-definite covariance matrix of the current Gaussian. The Mahalanobis distance has the unique property of being scale-invariant, making it effective for use on a 3DGS scene with a range of Gaussian sizes. Points with a Mahalanobis distance exceeding a predefined maximum threshold, which we set to a default sigma value of 2, are discarded.\newline
\begin{figure*}[t]
\centering
    \includegraphics[width=0.9\textwidth]{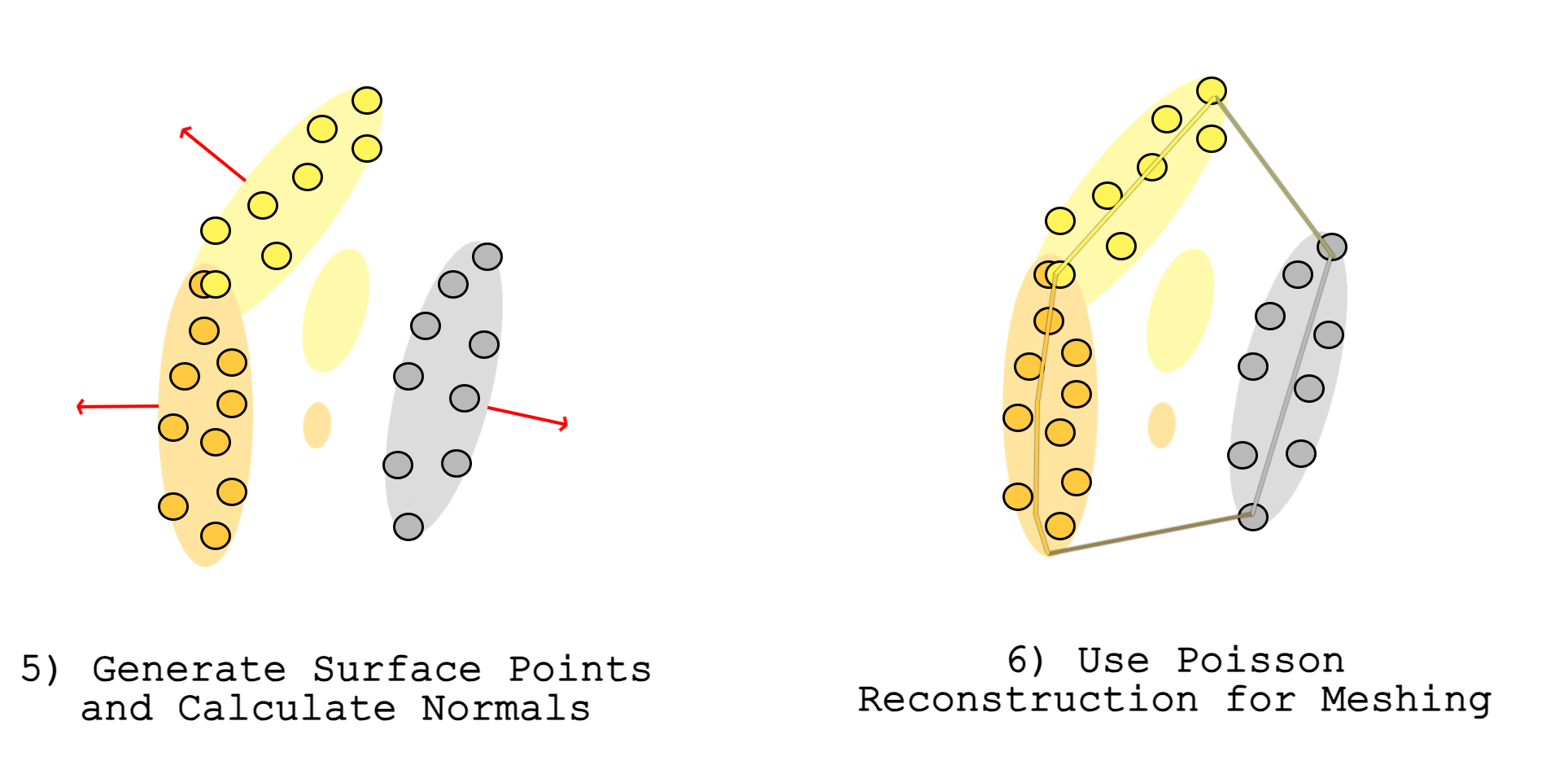}
     \caption{Demonstration of how this framework generates meshes from a point cloud. The process begins with identifying surface Gaussians using the Gaussian renderer, selecting only those with the highest contribution in each rendered image. Points are then sampled from these surface Gaussians to represent the scene's surfaces. The point normals are then calculated by taking the perpendicular vector of surface on the smallest side of each Gaussian. Finally, Poisson Surface Reconstruction connects the sampled points to generate the final mesh.}
    \label{fig:meshingShowcase}
\end{figure*}
For each Gaussian, resampling is performed for any points removed by thresholding until the required number of points is achieved. To maintain efficiency this resampling process is repeated up to a maximum of five times before proceeding to the next set of Gaussians with a different point allocation. This approach ensures that the final set of points accurately represents the Gaussian distributions, while also being efficient.  point sampling process is illustrated in Figure \ref{fig:samplingPointsShowcase}. \newline

\subsection{Meshing}

\begin{figure*}[t]
\centering
    \includegraphics[width=0.95\textwidth]{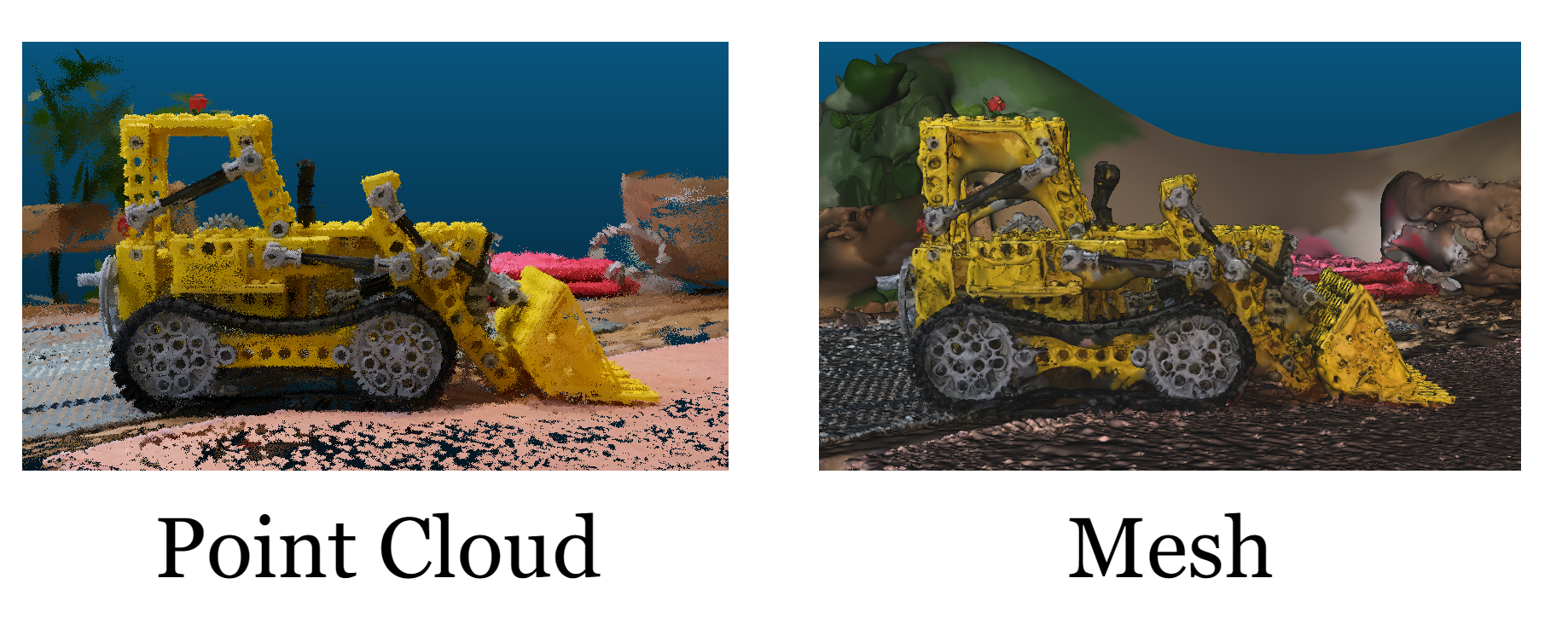}
     \caption{Closer comparison between a point cloud and mesh of the Mip-NeRF 360 dataset kitchen scene, generated using 3DGS-to-PC. The point cloud and meshes images were generated from the CloudCompare viewer.}
    \label{fig:MeshComparison}
\end{figure*}

In addition to generating a point cloud, our framework provides functionality to convert a 3DGS scene into a high-quality mesh. Converting a point cloud into a mesh is a challenging task, as points suspended in 3D space must be accurately assigned to surfaces to produce a series of faces with appropriate texturing and smoothing. This issue has been heavily researched, with techniques such as Poisson Surface Reconstruction \cite{kazhdan2006poisson} achieving notable success in converting dense point clouds into accurate meshes. \newline
These reconstruction methods rely on the availability of normals for each point to determine the orientation of the generated surface. We offer functionality for calculating normals using the same method as described in SuGAR \cite{guédon2023sugarsurfacealignedgaussiansplatting}. This process assigns the normal to the smallest side of each Gaussian. This approach is effective since we observe that Gaussians often elongate along surfaces they represent. \newline
Our generated point clouds contain points drawn from every Gaussian in the scene, not just those on object surfaces. Applying surface meshing algorithms such as Poisson Surface Reconstruction directly to such point clouds produces inaccurate results. Unlike traditional 3D representations, surfaces in 3DGS scenes are not immediately apparent and only become discernible when rendered. We therefore identify surface Gaussians by leveraging the same process used for colour calculation, as described in Section \ref{sec:rendering_colours}. Following the colour rendering process, each Gaussian has been assigned a maximum contribution value, representing the pixel colour to which it most contributed, and the proportion of this contribution. Surface Gaussians are determined by removing Gaussians with a maximum contribution lower than the mean of all Gaussian contribution values. Gaussians that are heavily occluded by other surfaces are therefore removed. This also implies the removal of heavily translucent Gaussians, which we find typically represent noise. We experimented with with other methods, such as only including Gaussians that contributed the most per pixel, but we found that this method produced the best results for meshing.\newline
We then generate a new point cloud, which is independent of the number of points specified by the user, using the filtered set of surface Gaussians. This refined point cloud is then cleaned using Open3D's \cite{Zhou2018} statistical outlier removal algorithm, to ensure noisy points are not included in the meshing process. The cleaned point cloud is then converted into a mesh using Poisson Surface Reconstruction \cite{kazhdan2006poisson}. To further enhance the mesh quality, Laplacian Smoothing \cite{vollmer1999improved} can then be optionally applied to reduce surface noise. The meshing process is illustrated in Figure \ref{fig:meshingShowcase}. A visual comparison between a series of 3DGS scenes, generated point clouds and meshes are shown in Figure \ref{fig:MipComparison}. A closer visual comparison between a point cloud and mesh generated on the Mip-NeRF 360 dataset kitchen scene is shown in Figure \ref{fig:MeshComparison}.

\begingroup
\setlength{\tabcolsep}{2.0pt}

\begin{figure*}
  \centering
  \bgroup
\def\arraystretch{1.5}%
  \begin{tabular}{cccc} 
  & \large \textbf{3DGS} \vspace{0.4em} & \large \textbf{Point Cloud} \vspace{0.4em}  & \large \textbf{Mesh} \vspace{0.4em}  \\
  \rotatebox[x=-1cm]{90}{\large Bike} \enspace \enspace & \includegraphics[width=5cm]{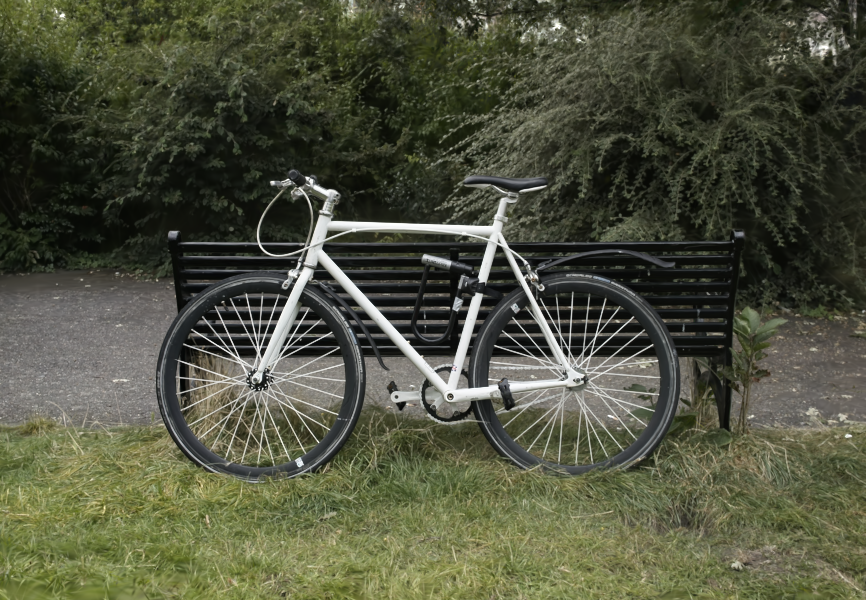} \enspace & \includegraphics[width=5cm]{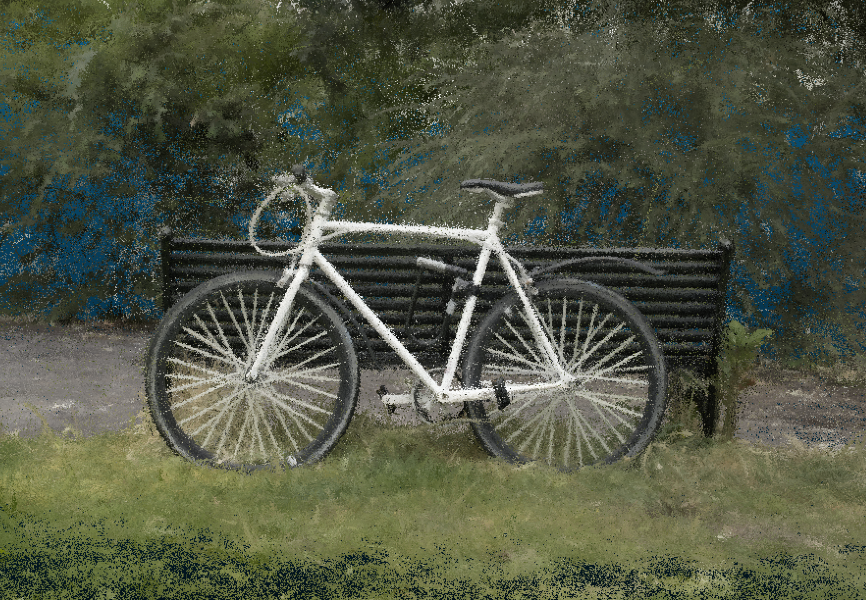} \enspace & \includegraphics[width=5cm]{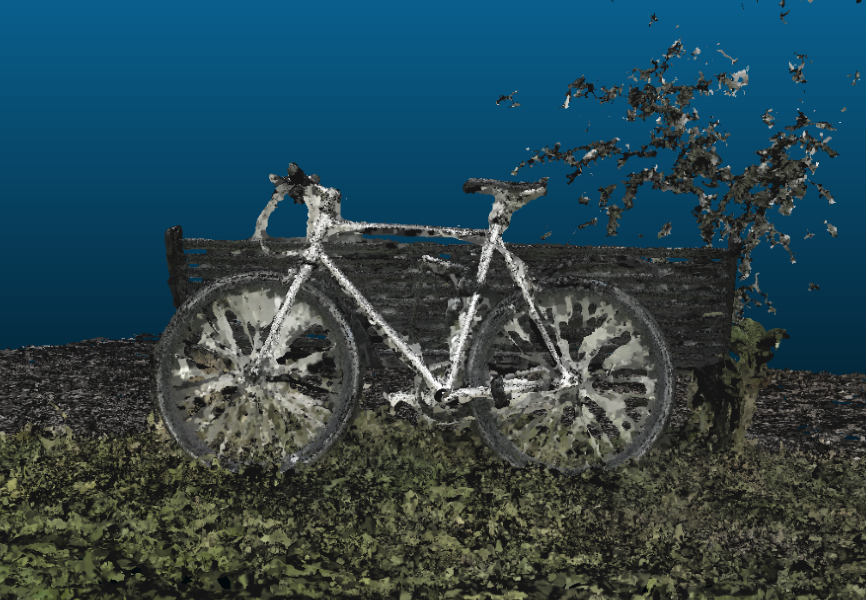}  \vspace{0.75em}   \\
\rotatebox[x=-1cm]{90}{\large Garden} \enspace \enspace & \includegraphics[width=5cm]{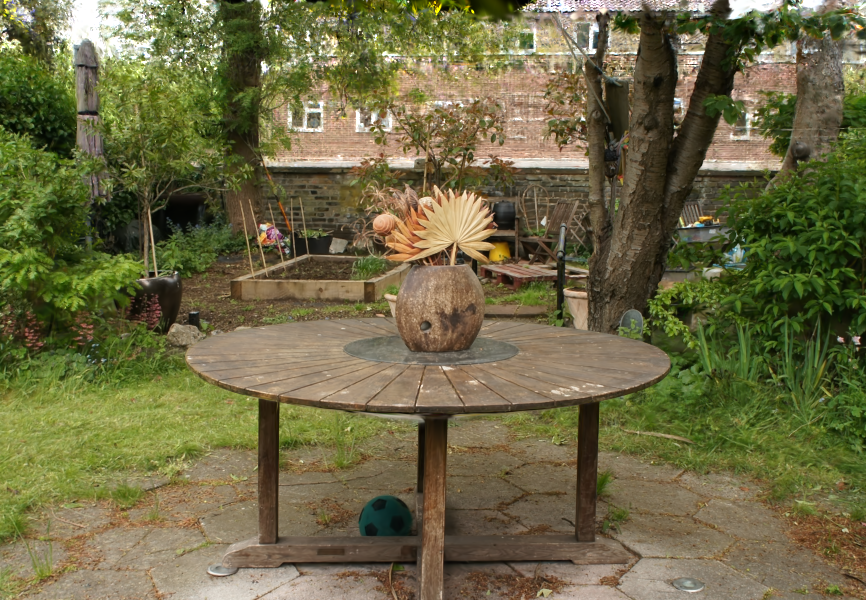} \enspace & \includegraphics[width=5cm]{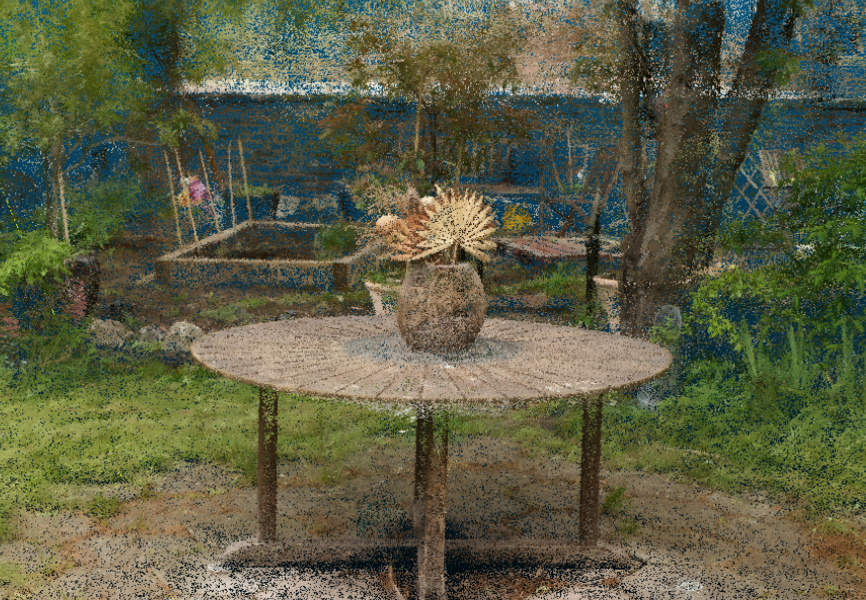} \enspace & \includegraphics[width=5cm]{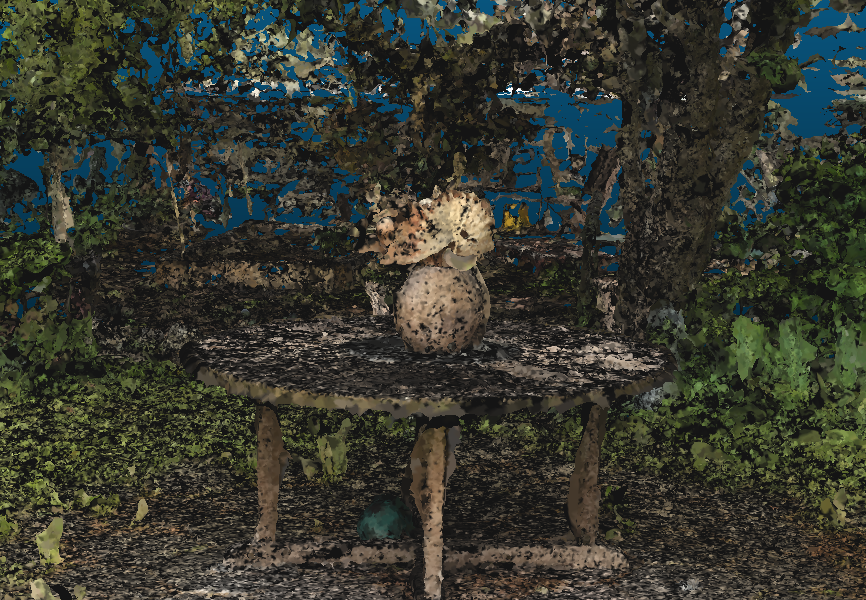} \vspace{0.75em} \\
  \rotatebox[x=-1cm]{90}{\large Stump} \enspace \enspace & \includegraphics[width=5cm]{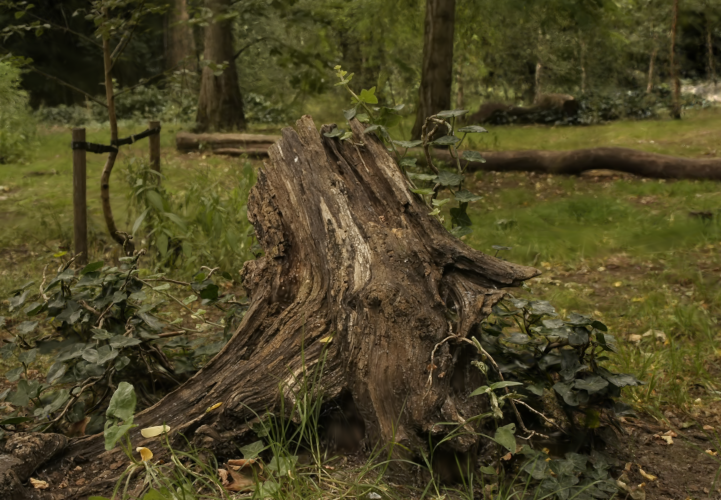} \enspace & \includegraphics[width=5cm]{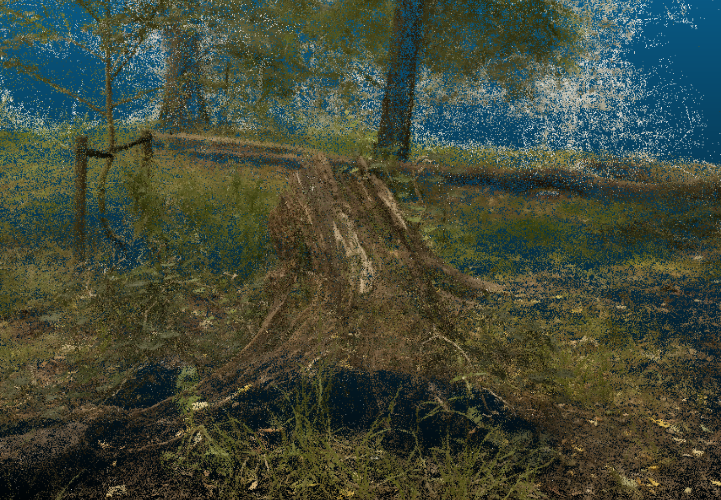} \enspace & \includegraphics[width=5cm]{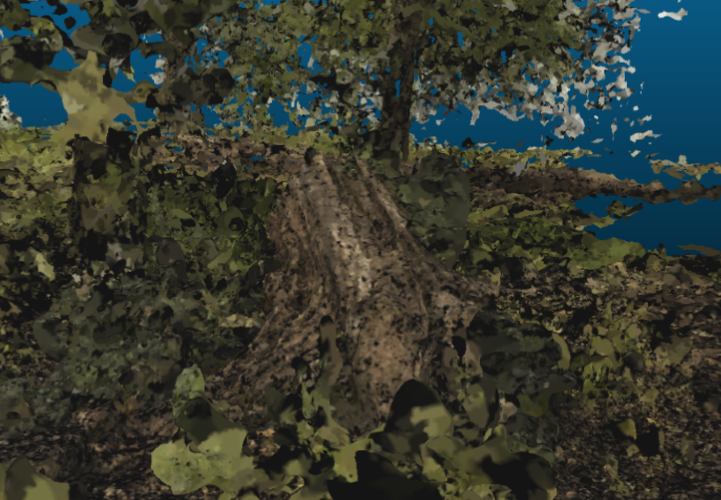} \vspace{0.75em} \\
 \rotatebox[x=-1cm]{90}{\large Kitchen} \enspace \enspace & \includegraphics[width=5cm]{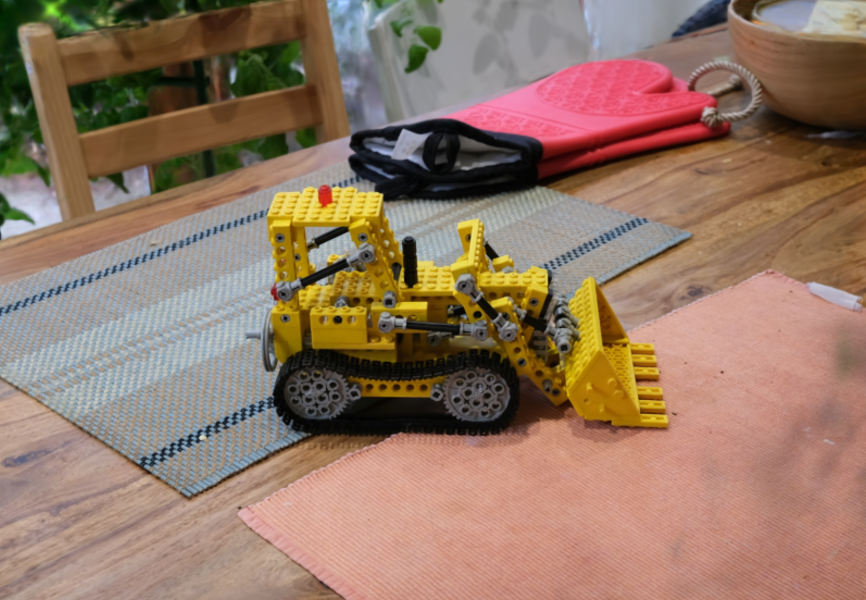} \enspace & \includegraphics[width=5cm]{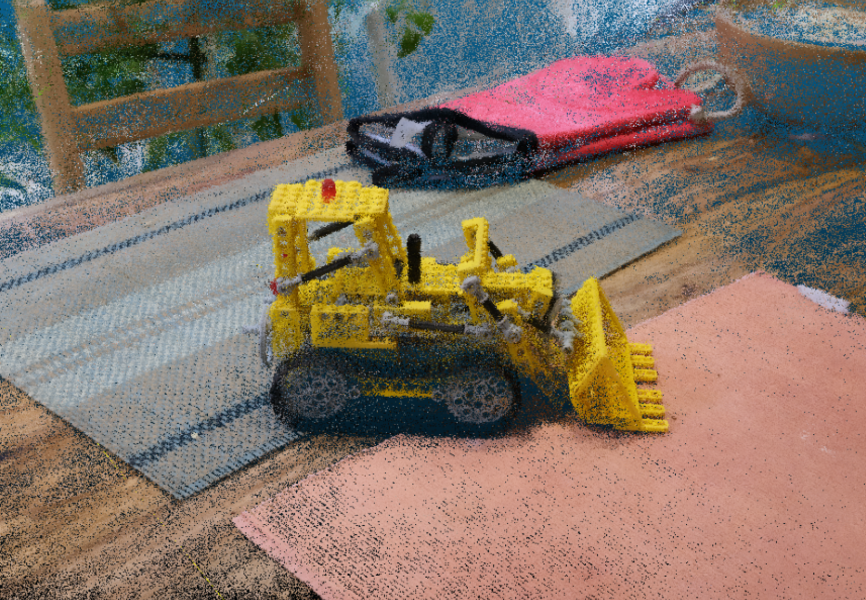} \enspace & \includegraphics[width=5cm]{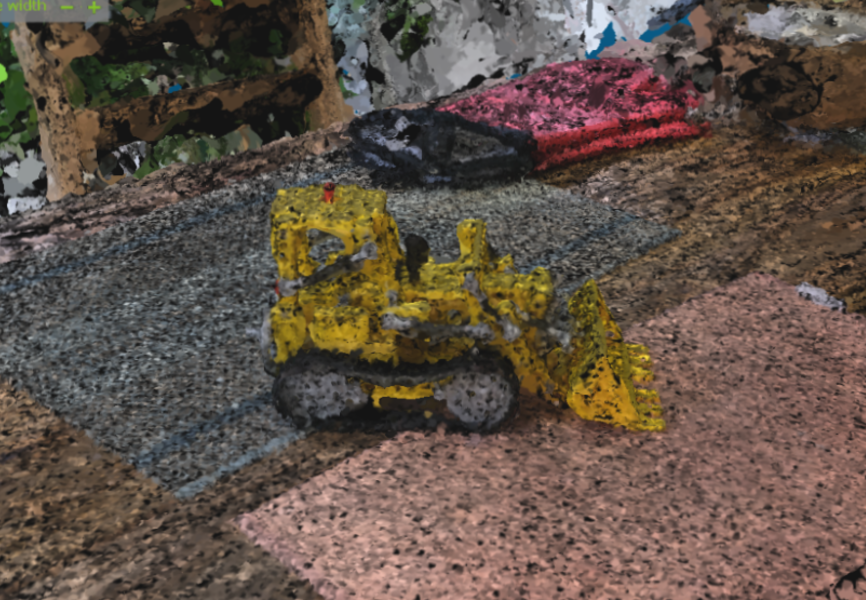} \vspace{0.75em} \\
  \end{tabular}
  \egroup
  \caption{Comparison between the 3DGS, point cloud and mesh representations for four different Mip-NeRF 360 dataset scenes. The point clouds and meshes were generated using 3DGS-to-PC on default arguments, apart from 'clean\_pointcloud' was set to true and the 'poisson\_depth' was set to 12 (to increase mesh quality). The Bike scene had an bounding box, -5 to 5 in every dimension, implemented only on the mesh, to improve visual clarity of the bike in the image. The 3DGS images were rendered using the official 3D Gaussian Splatting SIBR interactive viewer, while the point cloud and meshes images were generated from the CloudCompare viewer.}
  \label{fig:MipComparison}

\end{figure*}

\endgroup

\section{Discussion}

The 3DGS-to-PC framework is capable of generating highly dense and accurate point clouds from a range of 3DGS scenes. Its extensive customisation options allow users to reduce noise, adjust the number of points in the scene, and remove unwanted Gaussians based on specific characteristics. With the exception of Open3D, which is used for the point cloud to mesh conversion, all other packages utilised in the framework are consistent with those in the original 3DGS implementation. This compatibility enables straightforward intergration of the framework into existing 3DGS pipelines with minimal modification. \newline
A traditional method for generating point clouds involves converting an existing mesh into a point cloud. Tools such as CloudCompare can execute this process with ease by sampling points along the mesh surface. There exists a series of models that can produce meshes alongside a 3DGS scene, such as SuGAR \cite{guédon2023sugarsurfacealignedgaussiansplatting} and GaMeS \cite{waczyńska2024gamesmeshbasedadaptingmodification}. Hence, a valid approach to generate a point cloud of a 3DGS scene is to generate a mesh using these models, and then convert this into point cloud. However,
while these models are effective at generating high-quality meshes, they only represent the surfaces of the environment. Therefore, converting these meshes into point clouds results in points located solely on the mesh surfaces. \newline
In contrast, our approach generates points for every Gaussian in the scene, irrespective of its position in the environment. This results in a denser and more comprehensive representation of the scene. However, this approach does have the potential to introduce noise, since erroneous Gaussians that do not contribute to the scene surfaces are included. Despite this, incorporating all Gaussians in this process ensures a more authentic recreation of the original scene, making this framework particularly useful for applications requiring dense scenes with a high point count, or for comparisons against other points clouds. \newline
One limitation of the current implementation is the reliance on an altered version of the Torch-Splatting renderer. This renderer is written in pure Python and utilises PyTorch for handling Gaussians, which are represented as tensors. As a result, it is slower than the original 3DGS CUDA implementation, which benefits from direct GPU hardware access and low memory overhead when compared to Python. For example, in the Mip-NeRF 360 dataset, the average render time using the Python-based Gaussian renderer is approximately 2 seconds per frame, when executed on a Geforce RTX 2080 Ti graphics card with default argument values. Therefore, this process can become a bottleneck for scenes with a large number of camera poses. \newline
To address this issue, several options are avaliable to improve rendering speeds. First, the resolution of the rendered images can be reduced, which decreases the number of pixels involved in the colour calculations,  speeding up the rendering process. However, this may impact the final colour quality of the point cloud. Another approach is to render fewer images of the scene by skipping a percentage of cameras in the provided transforms. This is efficient for well-recorded scenes with cameras arranged in a linear path with overlapping images, which ensures that Gaussians are not accidentally omitted. \newline
Another limitation of the framework is that the dataset transforms are required to correctly render scene images. Dynamically selecting camera positions for rendering is challenging, as viewing the Gaussians from unanticipated angles and positions can result in incorrect final colours. Consequently, the original camera poses used to train the 3DGS scene are essential for accurately determining the new point colours. While most 3DGS datasets include the camera positions, this data is not universally available, posing a potential challenge for users working with only the trained 3DGS file. However, we still provide functionality to generate a point cloud if these transforms are not available. In this case, the point colours will be based on the original Gaussians, which are less accurate than the colours produced by our rendering process. \newline
Our meshing approach involves generating a point cloud of the predicted surfaces of the scene, and then using Poisson Surface Reconstruction to produce the final mesh. Although this method yields effective results, challenges remain in converting the point cloud into a mesh. The use of a multivariate normal distribution for point generation can cause noise in the surfaces, particularly for larger Gaussians. To address this, we recommend restricting the scene to a region of interest using a bounding box to improve meshing quality. \newline 
While methods such as SuGAR \cite{guédon2023sugarsurfacealignedgaussiansplatting} and GaMeS \cite{waczyńska2024gamesmeshbasedadaptingmodification} achieve more seamless meshing results, they require retraining the Gaussian representation of the scene. In contrast, our approach is compatible with any 3DGS scenes from any model that can produce a valid .ply or .splat file, offering greater flexibility and compatibility. Future improvements could involve incorporating algorithms to smooth the surface Gaussians prior to point cloud extraction. This enhancement would help reduce noise and improve the quality of the generated meshes.

\section{Conclusion}

In this work, we introduced 3DGS-to-PC, a robust framework for generating high-quality point cloud representations from 3DGS scenes. Our approach effectively calculates Gaussian colours by analysing their contributions to pixel colours in rendered images, ensuring accurate colour representation in the resulting point cloud. Points are distributed proportionally to each Gaussian’s volume. Outliers, which are identified via the Mahalanobis distance, are removed and regenerated to ensure authentic representation of the 3DGS scene. The framework also supports mesh generation through Poisson Surface Reconstruction applied to points sampled from predicted surface Gaussians. \newline
The framework is highly customisable, offering options for noise reduction, point cloud density control and Gaussian filtering. Furthermore, since most packages are shared with the original 3DGS implementation, this framework can be integrated into 3DGS pipelines. Future development will focus on reducing render times, and avoiding reliance on pre-existing camera poses for rendering new images.

\bibliographystyle{plain}
\bibliography{references}  

\clearpage

\end{document}